\documentclass{amsart}
\usepackage{graphicx}
\usepackage{epsfig}
\title {Tree Automata and Separable Sets of Input Variables }
\author{Sl. Shtrakov, Vl. Shtrakov}
\address{Dept. of Computer  Sciences, South-West University,
 Blagoevgrad  and  Dept. of Computer  Sciences, University of Sofia}
\email{shtrakov@aix.swu.bg }
\date{}
\newtheorem{l00}{\bf Lemma}

\newtheorem{t00}{\bf Theorem}
\newtheorem{e00}{\bf Example}

\newtheorem{d00}{\bf Definition}[section]

\def\Pr{\bf Proof.\ }
\def\fbx{\hfill${}^{\rule{2mm}{2mm}}$}
\def\A{{\mathcal A}}
\def\F{{\mathcal F}}
\def\la{\leftarrow}
\def\ra{\rightarrow}

\begin{document}
\maketitle
\begin{abstract}
We  introduce the separable sets of variables for trees and tree
automata. If a set $Y$ of input variables is inseparable for a
tree and an automaton then there a non empty family of
distributive sets of $Y$. It is shown that if a tree $t$ has
"many" inseparable sets with respect to a tree automaton $\A$ then
there is an effective way to reduce the complexity of $\A$ when
running on $t.$

\end{abstract}

\section{Introduction}
The consideration that finite automata may be viewed as unary
algebras is attributed to J.B\"uchi and J.Wright \cite{Ta}. In
many papers trees were defined as terms. Investigations on regular
and context-free tree grammars dated back to the  60-th.
\\
Tree automata are designed in the context of circuit verification and
logic programming. Since the end of 70's tree automata have been
used as powerful tools in program verification. There are many
results connecting properties of programs or type systems or
rewrite systems with automata \cite{Com,Ges}.
\\
The algebraic theory of terms was created and developed upto the
equational theory in the work of A.Malc'ev, G.Gr\"atzer
etc.\cite{Bu,Mal,Gr}.
\\
The theory of essential variables and separable sets for discrete
functions was created and developed by S.Jablonsky, A.Salomaa,
K.Chimev etc.\cite{Ch,Ja,Sa}.  The results obtained here are very
useful for analysis and synthesis of functional schemes and
circuits.
\\
The present paper is a continuation and generalization of the
results in \cite{[Dr2000]} which  are  borderline cases of  these
 fields of theoretical computer science and mathematics.

\section{Preliminaries}
Let ${\F}$ be any finite set, the elements of which are called $
operation\ symbols. $ Let $\tau:{{\F}}\to N$ be a mapping into the
non negative integers; for $f\in{\F},$ the number $\tau(f)$ will
denote the {\it arity } of the operation symbol $f.$ The pair
$(\F,\tau)$ is called {\it type} or {\it signature}. If it is
obvious what the set ${\F}$ is, we will write "$ type\ \tau$". The
set of symbols of arity $p$ is denoted by ${\F}_p.$ Elements of
arity $0,1,\ldots, p$ respectively are called {\it
constants(nullary), unary,...,$p$-ary} symbols. We assume that
${\F}_0\neq\emptyset.$

\begin{d00}\rm\label{d01}
Let $X=\{x_1,x_2\ldots,\}$ be a set of distinct objects called
variables, and let $\tau$ be a type with the set of operation
symbols ${\mathcal F}=\cup_{i\geq 0}{\mathcal F}_i=(f_i)_{i\in
I}.$ The set $W_{\tau}(X)$ of {\it terms\ of\ type\ $\tau$ } with\
variables\ from\ $X $ is the smallest set such that
\\
$(i)$\ $X\cup\F_0\subseteq W_\tau(X);$
\\
$(ii)$\ if $f $ is $n-$ary operation symbol and $t_1,\ldots,t_{n}$
are terms then the "string" $f(t_1\ldots t_{n})$ is a term.
\end{d00}
Note that terms are also called {\it trees.}
\\
Let $t$ be a term then the set $Var(t)$ consisting of these
elements of $X$ which occur in $t$ is called the set of
 {\it input variables (or variables)} for this term.
 \\
  The $depth$ of a tree $t$ is defined
in the  following inductive way:\\
$(i)$ If $t\in X\cup \F_0$ then $Depth(t)=0;$\\
$(ii)$ If $t=f(t_1,\ldots,t_{n})$ then
$Depth(t)=max \{Depth(t_1),\ldots,Depth(t_{n})\} +1.$
\\
If $t=f(t_1,\ldots,t_n)$ then $t,\ t_1,\ldots,t_n$ are {\it
subterms (subtrees)} of $t$ and all subtrees of $t_1,\ldots,t_n$
are subtrees of $t,$ too.
\\
Thus we define a partial order relation in the set of all terms
$W_{\tau}(X).$ We denote by $\unlhd$ the subterm ordering, i.e. we
write $t\unlhd t'$ if $t$ is a  subterm of $t'.$ We denote $t\lhd
t'$ if $t\unlhd t'$ and $t\neq t'.$ A chain of subterms $t_1\lhd
t_2\lhd \ldots \lhd t_k$ is called {\it strong} if there does not
exist a term $s$ such that $t_j\lhd s\lhd  t_{j+1}$ for some $j\in
\{1,\ldots,k-1\}.$
\\
Let $t,t'\in W_{\tau}(X)$ and $t_1\unlhd t.$ We  denote by
$t(t_1\la t')$ the term which is obtained by substituting in $t$
simultaneously $t'$ for each occurrence of $t_1$ as a subterm of
$t.$
\section{Finite Tree Automata  and
Separable Sets of  Input Variables}
\begin{d00}\rm\label{d5}
A  {\it finite tree automaton } over $\F$ and $X$ is a tuple
$\A=\langle Q,\F, X, Q_f,\Delta\rangle$ where, $\F$ and $X$ are
sets of operational symbols and variables,
 $Q$ is a finite set of states,
\
 $Q_f\subseteq Q$ is a set of final states and  $\Delta$ is the set
 of transition rules,
 $ \Delta=\{\Delta_0,\Delta_1,\ldots,\Delta_n\},$ where $\Delta_0:\F_0\ra Q,$
and
$\Delta_i:\F_i\times Q^i\ra Q, \ i=1,\ldots,n$  are mappings.
In this paper we will consider  complete and deterministic automata only i.e. $\Delta_i$ is a total function
 for each $ i=0,1,\ldots,n.$
\end{d00}

Let $Y\subseteq X$ be a set of variables and $\gamma:Y\ra \F_0$ be
a function which assigns  nullary operation symbols (constants) to
each input variable from $Y.$
 The function $\gamma$ is called
{\it assignment} on the set of inputs $Y.$ The set of such assignments
 will be denoted by
$Ass(Y,\F_0).$

Let $t\in W_{\tau}(X),$ $\gamma\in Ass(Y,\F_0)$ and
$Y=\{x_1,\ldots,x_m\}.$ The term $t(x_1\la
\gamma(x_1),\ldots,x_m\la \gamma(x_m))$ will be denoted by
$\gamma(t).$ \ We will definitely assume that if $x_i\in
Y\setminus Var(t)$ then $t(x_i\la\gamma(x_i))=t$ for each
$\gamma\in Ass(Y,\F_0).$

It is clear that if \label{l7}  $Y\cap Z=\emptyset$ , $\gamma_1\in
Ass(Y,\F_0)$ and $\gamma_2\in Ass(Z,\F_0)$ then
$\gamma_1(\gamma_2(t))= \gamma_2(\gamma_1(t)).$

Let $\gamma\in Ass(X,\F_0)$. The automaton  $\A=\langle Q,\F, X,
Q_f,\Delta\rangle$
  runs on $t$ and $\gamma.$ It starts at
leaves of $t$ and moves downwards, associating along the run a
resulting state with each subterm inductively:
\\
$(i)$\quad If $Depth(t)=0$ then the automaton $\A$ associates the state $q\in Q$ with
$t$,  where $q=\Delta_0(\gamma(x_i))$ if
$t=x_i\in X$ and $q=\Delta_0(f_0)$ if $t=f_0\in \F_0.$
\\
$(ii)$\quad Let $Depth(t)\geq 1.$ If $t=f(t_1,\ldots,t_n)$ and the
states  $q_1,\ldots,q_n$ are associated with the subterms
$t_1,\ldots,t_n$ then  the automaton $\A$ associates the
state $q$ with $t$, where $q=\Delta_n(f,q_1,\ldots,q_n).$
\\
A term $t$ in $W_{\tau}(X)$ is accepted by an automaton
$\A=\langle Q,\F, X, Q_f,\Delta\rangle$ if there exists an
assignment $\gamma$ such that when running on $t$ and $\gamma$
the automaton $\A$ associates with $t$ a final state  $q\in Q_f.$
\vspace{.5cm}

 When $\A$ associates the state $q$ with a tree $s,$ and an assignment $\gamma\in Ass(X,\F_0)$
 we will write $ \A(\gamma,s)=q.$
\begin{d00}\rm\label{d6}
An input variable $x_i\in Var(t)$ is called {\it essential} for
$t$ and $\A$ if there exist  two assignments $\gamma_1,
\gamma_2\in Ass(X,\F_0)$ such that $\gamma_1(x_j)= \gamma_2(x_j),$
for each variable $x_j, x_j\neq x_i$  and $\A(\gamma_1,t)\neq
\A(\gamma_2,t)$.
\end{d00}
The set of all essential inputs for $t$ and $\A$ is denoted by
$Ess(t,\A).$ The input variables from
$Var(t)\setminus Ess(t,\A)$
are called {\it fictive } for  $t$ and $\A.$
\begin{l00}\rm\label{l10}
Let $f_0\in\F_0.$ If $x_i\notin Ess(t,\A)$ then
\begin{center}
$\A(\gamma,t)=\A(\gamma,t(x_i\la f_0))$
\end{center}
for each $\gamma\in Ass(X,\F_0).$
\end{l00}
\Pr\rm Suppose the lemma is false and let $\gamma_0\in
Ass(X,\F_0)$ be an assignment such that
$\A(\gamma_0,t)\neq\A(\gamma_0,t(x_i\la f_0)).$ \ Consider the
assignment $\gamma_1\in Ass(X,\F_0)$ defined by $\gamma_1(x)= f_0$
if $x=x_i,$ and $\gamma_1(x)=\gamma_0(x)$ if $x\neq x_i.$ Hence
$\A(\gamma_1,t)=\A(\gamma_0,t(x_i\la f_0))\neq\A(\gamma_0,t),$
i.e. $x_i\in Ess(t,\A).$ A contradiction.
\fbx
\begin{l00}\rm\label{l11}
Let $t,s\in W_{\tau}(X).$ If $x_i\notin Ess(t,\A)$
and for each $ q\in Q$ there exists $f_0\in \F_0$ such that
$\Delta_0(f_0)=q$
then
\begin{center}
$\A(\gamma,t)=\A(\gamma,t(x_i\la s))$
\end{center}
for each $\gamma\in Ass(X,\F_0).$
\end{l00}
\Pr\rm Suppose that the lemma is false and let $\gamma_0\in
Ass(X,\F_0)$ be such assignment that
$\A(\gamma_0,t)\neq\A(\gamma_0,t(x_i\la s)).$ \ Since $t(x_i\la
s)\in W_{\tau}(X)$ and $\A$ is complete, it follows that there is
a state $q,\ q\in Q$ such that $\A(\gamma_0,s)=q.$ Let
$f_0\in\F_0$ be such nullary operation symbol that
$\Delta_0(f_0)=q.$ \ Hence $\A(\gamma_0,t(x_i\la
s))=\A(\gamma_0,t(x_i\la f_0)).$ Now, as in  Lemma \ref{l10} we
will obtain $x_i\in Ess(t,\A)$ which is a contradiction. \fbx
\begin{d00}\rm\label{d13}
 A set
$Y\subseteq Ess(t,\A)$ is called {\it separable} for $t$ and $\A$
w.r.t. a set $Z\subseteq Ess(t,\A),$ with $ Z\cap Y=\emptyset$ if
there is an assignment $\gamma$ on $Z$ such that $Y\subseteq
Ess(\gamma(t),\A).$
\end{d00}
 The set of all separable sets for $t$ and $\A$ w.r.t. $Z$ will be denoted
by $Sep(t,\A,Z).$ When $Y$ is separable for $t$ and $\A$ w.r.t.
$Z=Ess(t,\A)\setminus Y$ the set $Y$ is called {\it separable} for $t$ and
$\A$ and the set of such $Y$ will be denoted by
$Sep(t,\A).$
\\
When a set of essential inputs is not separable, it will be called {\it inseparable}.
\begin{t00}\label{t1}\rm
If $Y\in Sep(t,\A)$ then
for every input $x_i\in Y$
there exists at least one strong
chain
$x_i=t_k\lhd t_{k-1}\lhd \ldots \lhd t_1\unlhd t$
such that $x_i\in Ess(t_j,\A)$ for $j=1,\ldots,k.$
\end{t00}
The proof of the theorem can be done as Theorem 1 in \cite{[Dr2000]}.
\begin{t00}\label{l4}\rm
If $\A(\gamma,t_1)=\A(\gamma,t)$ for every $ \gamma\in
Ass(X,\F_0)$  then $Sep(t,\A)=Sep(t_1,\A).$
\end{t00}
\Pr\rm Let $Y\in Sep(t,\A)$ and $Y=\{x_1,\ldots,x_m\}.$ There is
an assignment $\gamma_0\in Ass(Z,\F_0),$ $ Z=X\setminus Y,$ such
that $Y=Ess(\gamma_0(t),\A).$ We have to prove that $Y\subseteq
Ess(\gamma_0(t_1),\A).$ \ Let $x_i\in Y$ be an arbitrary input
variable from $Y.$
 It follows that there are two assignments
$\gamma_1,\gamma_2\in Ass(X,\F_0)$ with
$$\forall x_j\notin Y \quad \gamma_1(x_j)=\gamma_2(x_j)=\gamma_0(x_j),\quad
 \forall x_j\in Y,\ j\neq i \quad \gamma_1(x_j)=\gamma_2(x_j)$$
 and
$(\gamma_1(x_i)\neq\gamma_2(x_i)$ such that $\A(\gamma_1,t)\neq
\A(\gamma_2,t).$ Hence $\A(\gamma_1,t_1)= \A(\gamma_1,t)\neq
\A(\gamma_2,t)= \A(\gamma_2,t_1)$ i.e. $x_i\in
Ess(\gamma_0(t_1),\A).$ Consequently $Sep(t,\A)\subseteq
Sep(t_1,\A).$
\
The inclusion $Sep(t_1,\A)\subseteq Sep(t,\A)$
can be proved in a similar way.
\fbx
\\
The following lemma is obvious.
\begin{l00}\rm\label{l21}
If $Y\notin Sep(t,\A,Z)$
and $V\subset Ess(t,\A)$ with $V\cap Z=\emptyset$
then
$Y\cup V\notin
Sep(t,\A,Z).$
\end{l00}

Further, we want to describe what  the relation between separable
sets for $t$ and $\A$ and the "speed of runs" of $\A$ on $t$ is?

Let us consider the following two transformations of $t$,
depending on $\A$:
\\
$(i)$\ if $x_i$ is fictive for $t$ and $\A$ and $f_0\in\F_0$ then
as result we obtain the tree  $t'=t(x_i\la f_0);$
\\
$(ii)$\ if $t_1\lhd t_2\unlhd t$ with
$\A(\gamma,t_1)=\A(\gamma,t_2)$ for each assignment $\gamma\in
Ass(X,\F_0)$ then as result we have $t'=t(t_2\la t_1).$

When $t'$ is an image of $t$ under such a transformation we will
write
 $t\vdash_\A t'.$
The transitive closure  of $\vdash_\A $ in $W_\tau(X)$ will be
denoted by $\models_\A.$

 \begin{t00}\rm \label{l20}
For every two terms $t$ and $s$ if
 $t\models_\A s$
then $\A(\gamma,t)=\A(\gamma,s)$ for every assignment $\gamma\in
Ass(X,\F_0).$
\end{t00}
\Pr\rm
 Let $t\vdash_\A s.$ \
 If $Dept(t)=0$ then
$t=x_i$ or $t=f_0$ for some $f_0\in\F_0.$ Clearly $s=t$ and the
theorem is proved in this case. \ Let $Depth(t)\geq 1.$ At first
let $s$ be  a  term obtained through applying a transformation
with $t_2\in X.$ Hence
 \
$t=f(t_1,\ldots, t_n),$
 with
$x_i\notin Ess(t,\A).$ Let $t_{i_1},\ldots, t_{i_k}$ be all
subterms amongs $t_1,\ldots, t_n$ for which
 $x_i\in Var(t_{i_p}), \ p=1,\ldots,k.$
\ Then $s=f(t_1,\ldots,t'_{i_1},\ldots,t'_{i_k},\ldots,
t_n)=t(t_{i_1}\la t'_{i_1}, \ldots,t_{i_k}\la t'_{i_k})=t(x_i\la
f_0)$ where $t'_{i_p}=t_{i_p}(x_i\la f_0), \ p=1,\ldots,k$ for
some $f_0\in\F_0.$ \ Hence for all $\gamma_1,\gamma_2\in
Ass(X,\F_0)$ if $\gamma_1(x_j)=\gamma_2(x_j)$ with $j\neq i$ then
$\A(\gamma_1,t)=\A(\gamma_2,t).$ \ Let $\gamma\in Ass(X,\F_0)$ be
an arbitrary assignment and let us consider the assignment
$\gamma'\in Ass(X,\F_0)$ defined as follows: $\gamma'(x)=f_0 $ if
$ x=x_i$ and $\gamma'(x)=\gamma(x)$ if $x\neq x_i.$ \ Thus we have
$\A(\gamma',t)=\A(\gamma,t)$ and $\A(\gamma',t)=\A(\gamma,t(x_i\la
f_0))=\A(\gamma,s).$ The theorem is proved in this case.

Let $s$ be a  term obtained through applying a transformation with
$t_2,$ $ Depth(t_2)>0.$
 Hence there are subterms $t_1\lhd t_2\unlhd
t$ with $ \A(\gamma,t_1)=\A(\gamma,t_2)$ for every $\gamma\in
Ass(X,\F_0)$ and $s=t(t_2\la t_1).$ Clearly
$\A(\gamma,s)=\A(\gamma,t(t_2\la t_1))=\A(\gamma,t(t_2\la t_2))=
\A(\gamma,t).$
 \fbx

\section{Complexity of  Automata on Trees}

It is easy to see that if $t\lhd s$ with
$\A(\gamma,t)=\A(\gamma,s)$ for each assignment $\gamma\in
Ass(X,\F_0)$ then the results of the runs of $\A$ on $t$ and $s$ will be the same, but the run on
$t$ will be "quicker" than the run on $s$ because of $t\lhd s.$ So, we need a definition of the "quickness" of runs of an automaton on a tree.

Let $t$ be
a tree and $\A$ be an automaton. The set of all  states of
$\A$ which can be associated with $t$ will be denoted by
$St(t,\A)$ and $st(t,{\mathcal A})=|St(t,\A)|$ is the number of
the elements in $St(t,\A).$
Thus $q\in St(t,\A)$ if and only if there is an assignment $\gamma\in Ass(X,\F_0)$ such that  $\A(\gamma,t)=q.$
\begin{d00}\rm\label{d16}
The {\it complexity } of
$\A$ on  $t$ denoted by $Comp(t,\A)$ is defined in the following inductive way:
\\
$(i)$\quad If $t=x\in X$ then $Comp(t,\A)=\sum_{f_0\in\F_0}st(f_0,\A);$
\\ $(ii)$\quad
If $t=f_0\in \F_0$ then $Comp(t,\A)=st(f_0,\A);$
\\
$(iii)$\quad If $t=f(t_1,\ldots,t_n)$ then $$Comp(t,\A)=\prod_{j=1}^n st(t_j,\A)+
\sum_{i=1}^n Comp(t_i,\A).$$
\end{d00}
If  automaton $\A$ is complete and deterministic then
$Comp(x,\A)=|\F_0|$ and $Comp(f_0,\A)=1, \ f_0\in\F_0.$

So, the complexity  of $\A$ on $t$ presents the number of all calculations of values of $\Delta$
for all runs of $\A$ on $t.$

It is clear that if $t\models_\A s$ then  $Comp(s,\A)\leq Comp(t,\A)$.
\begin{e00}\label{e1}\rm
Let $\A=\langle Q,\F, X, Q_f,\Delta\rangle$  with \
$\F_0=\{0,1,2\}$, $\F_1=\{f_0,f_1,f_2\}$, $\F_2=\{g\},$
$\F_4=\{h\},$ \ $Q=\{q_0,q_1,q_2\}$, $Q_f=\{q_1\}$, \
$\Delta_0(i)=q_i$ for  $i=0,1,2$, \
$\Delta_1(f_i,q_j)=\left\{\begin{array}{ll}
       q_1, & \mbox{\rm if}\quad i=j\cr
       q_0,  & \mbox{\rm if} \quad i\neq j;
        \end{array}
        \right.
$
for  $i=0,1,2$,
\
$\Delta_2(g,q_i,q_j)=q_m,$ where $m=i.j\ (mod\ 3)$ and
\
$\Delta_4(g,q_i,q_j,q_k,q_l)=q_m,$ where $m=i+j+k+l\ (mod\ 3).$
\\
Let us consider the term
$t=h(g(f_0(x_1),x_2),g(f_1(x_1),x_3),g(f_2(x_1),x_4),x_5),$ with the
tree, given on the Figure \ref{f1}

\begin{figure}
  \includegraphics[width=10cm]{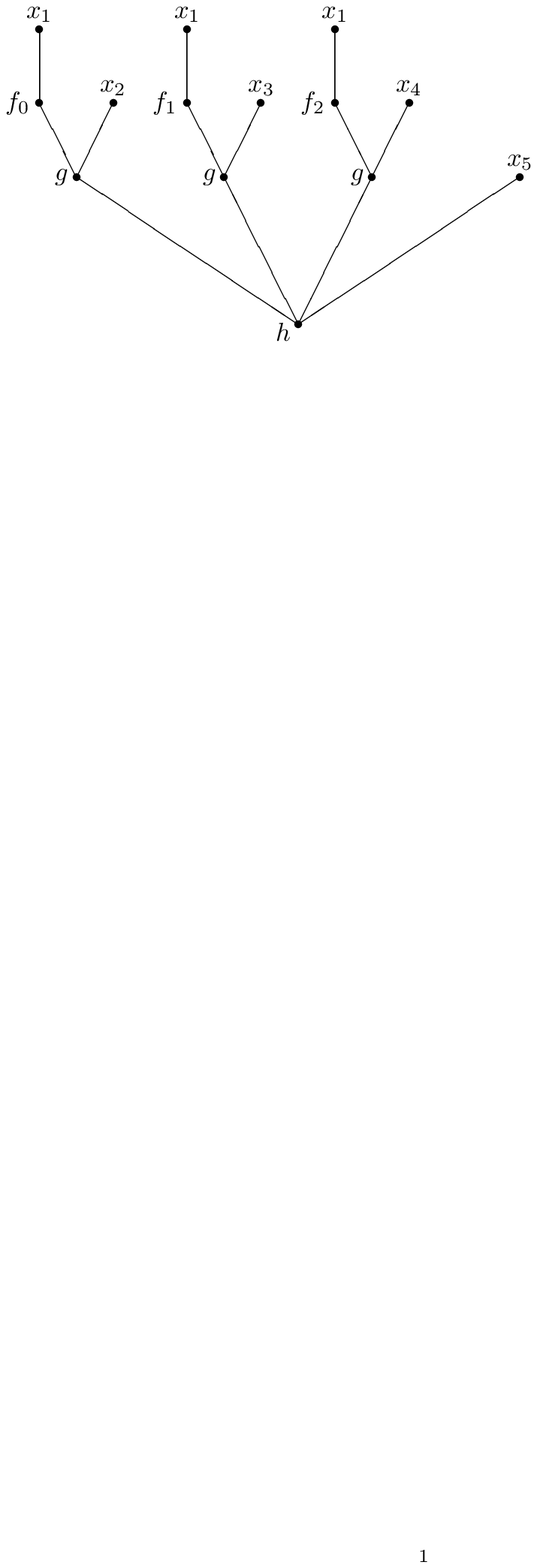}\\
  \caption{Tree of a term}\label{f1}
\end{figure}

The subterms of this term are:
\
$t_1=g(f_0(x_1),x_2),$
$t_2=g(f_1(x_1),x_3),$
$t_3=g(f_2(x_1),x_4),$
$t_4=x_5,$
\
$t_{11}=f_0(x_1),\ t_{12}=x_2,$
$t_{21}=f_1(x_1),\ t_{22}=x_3,$
$t_{31}=f_2(x_1),\ t_{32}=x_4,$
\
$t_{111}=x_1,\ t_{211}=x_1,\ t_{311}=x_1. $
\\
Let us calculate $Comp(t,\A).$ Clearly
\\
$Comp(t_{111},\A)=Comp(t_{211},\A)=
Comp(t_{311},\A)=Comp(t_{12},\A)=$
\\ $ = Comp(t_{22},\A)=
Comp(t_{32},\A)$ $=Comp(t_{4},\A)=3.$
 \ Because  $f_i\in\F_1$ and
$st(x_1,\A)=3$ it follows that $Comp(f_i(x_1),\A)=6$\ for
$i=0,1,2,$ i.e. \
$Comp(t_{11},\A)=Comp(t_{21},\A)=Comp(t_{31},\A)=6,$ \ Let us note
that $St(t_{i1},\A)=\{q_0,q_1\}$ for $i=1,2,3$ and
$st(t_{i1},\A)=2$ for $i=1,2,3.$ \ Analogously, $st(t_{i2},\A)=3$
for $i=1,2,3.$ Thus $Comp(t_{i},\A)=2.3+6+3=15$ for $i=1,2,3$ i.e.
$Comp(t_{1},\A)=Comp(t_{2},\A)=Comp(t_{3},\A)=15.$ \ It is easy to
see that $st(t_{i},\A)=3$ for $i=1,2,3,4.$ Hence
$Comp(t,\A)=3.3.3.3+15+15+15+3=129.$
\end{e00}
\section{Distributive Sets of Inseparable Sets of Inputs}
We will consider the case when a set of essential inputs is
inseparable. It seems that if a term has "many" inseparable sets
the runs of $\A$ on such a term will be  "quicker".
\begin{d00}\rm\label{d14}
Let $Y,Z\subseteq Ess(t,\A),\quad Y\cap Z=\emptyset$ and $Y\notin
Sep(t,\A).$ The set $Z$ is called {\it distributive set
 } of $Y$ for $t$ and $\A$ if $ Y\not\subseteq Ess(\gamma(t),\A)$ for every $\gamma\in
Ass(Z,\F_0)$
and $Z$ is
minimal with respect to this property.
\end{d00}
The family of all distributive sets  of $Y$ will be denoted by $Dis(Y,t,\A).$
Note that the family of distributive sets of $Y$ is non-empty iff $Y$ is not separable.

\begin{t00}\rm\label{t2}
If $Z\in Dis(Y,t,\A)$ then for each proper subsets $Z_1$ and $Y_1$
of $Z$ and $Y$ it is held that $Z_1\notin Dis(Y_1,t,\A).$
\end{t00}
\Pr\rm Let $Y_1$ is a proper subset of $Y.$ Suppose the theorem is
false and let $Z_1$ is a proper subset of $Z$ with $Z_1\in
Dis(Y_1,t,\A).$ Because of Lemma \ref{l21} it follows
 that $Z_1\in Dis(Y,t,\A).$ This contradicts to the minimality of $Z$ as a distributive set of $Y$ and $\A.$
 \fbx

The next example is a good illustration of how to use distributive
sets to obtain "quicker" runs of $\A$ on $t$ under different
assignments.
\begin{e00}\rm
Let us try to find a simpler way for running of $\A$ on $t$ and
$\gamma\in Ass(X,\F_0)$ where $t$ and $\A$ are as in Example
\ref{e1}.

 Let $Y=\{x_2,x_3,x_4\},$ $Z=\{x_1\}$ and
$\gamma\in Ass(Z,\F_0).$ There are only the following three possible cases.
\\
a)\ If $\gamma(x_1)=0$ then $x_3,x_4\notin Ess(\gamma(t),\A);$
\\
b)\ if $\gamma(x_1)=1$ then $x_2,x_4\notin Ess(\gamma(t),\A);$
\\
c)\ if $\gamma(x_1)=2$ then $x_2,x_3\notin Ess(\gamma(t),\A).$
\\
Hence $Y\notin Sep(t,\A)$ and
$Z\in Dis(Y,t,\A).$
\\
Now, we can consider $Comp(t,\A)$ and
use distributive set $Z$ to obtain simpler
runs of $\A$ on $t.$
\
 The fact that $Z$ is a distributive set of $Y$
allows us to
 distribute all 243 assignments in three classes $\Gamma_0,\Gamma_1,\Gamma_2$
according to a),b) and c) i.e. $\gamma\in \Gamma_i\iff \gamma(x_1)=i.$
\
Let $\gamma\in Ass(X,\F_0)\cap \Gamma_0.$
 We can apply a transformation defined as above on
the tree \ $\gamma(t)=
h(g(f_0(0),x_2),g(f_1(0),x_3),g(f_2(0),x_4),x_5).$ \ By
$\Delta_1(f_i,q_j)=0$ when $i\neq j$ it follows that
$\gamma(t)\models_{\A} s_0,$ where $s_0= h(x_2,0,0,x_5)$ (see Figure
\ref{f2}). It is easy to calculate $Comp(s_0,\A)=17.$ \ In an
analogous way the trees $s_i$ (see Figure \ref{f2}) when $\gamma\in
Ass(X,\F_0)\cap \Gamma_i,\    \ i=1,2$ with $Comp(s_i,\A)=17,   \
i=1,2$ can be obtained.
\begin{figure}
    \includegraphics[width=13cm]{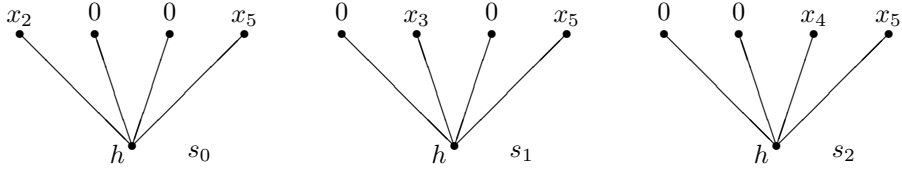}\\
  \caption{Distributed trees }\label{f2}
\end{figure}

So, we have a very simple procedure to execute the runs of $\A$
on $t$ with given $\gamma\in Ass(X,\F_0).$ This procedure consists of:
\\
Step 1. Find $i,\ i\in\{0,1,2\}$ such that $\gamma\in\Gamma_i.$
\\
Step 2. Find $\A(\gamma,s_i).$
\par
Note that  step 1. can be realized by a simple checking
$\gamma(x_1)=0|1|2.$ We can naturally assume that the
 complexity of this step equals  $3.$ Thus the
 complexity of the whole procedure is 20 and in the general
case it is 129.
\end{e00}

This example is a good motivation
for future investigations of the inseparable sets and
their distributive sets.
\begin{t00}\rm\label{t21}
If $Z\in Dis(Y,t,\A)$ then for each proper subsets $Z_1$ and $Y_1$
of $Z$ and $Y$ it is held that $Z_1\notin Dis(Y_1,t,\A).$
\end{t00}
\Pr\rm
Let $Y_1$ is a proper subset of $Y.$ Suppose the theorem is false and
let $Z_1$ is a proper subset of $Z$ with $Z_1\in Dis(Y_1,t,\A).$
Because of Lemma \ref{l21} it follows
 that $Z_1\in Dis(Y,t,\A).$ This is a contradiction with the
minimality of $Z$ as a distributor of $Y$ and $\A.$
\fbx
\begin{d00}\rm\label{d17}
Let ${\mathcal M}=\{M_1,\ldots,M_m\}$ be a finite family of
 nonempty sets. A set $M=\{z_1,\ldots,z_l\}$
is called {\it  representative system } for ${\mathcal M}$ if $
M\cap M_i\neq\emptyset$ for every $i\in\{1,\ldots,m\}$ and $M$ is
minimal with respect to  this property.
\end{d00}
\begin{l00}\rm\label{l8}
If $M$ is  a representative system for ${\mathcal M}$ then the
following is true:
\\
$(i)$ For each  $ M_i\in{\mathcal M}$ there is $ z_j\in M$ with
$z_j\in M_i;$
\\
$(ii)$ For each  $ z_j\in{M}$ there is $ M_i\in{\mathcal M}$ with
$\{z_j\}= M_i\cap M.$
\end{l00}
\Pr\rm The statement $(i)$ is obvious. To prove $(ii)$ let us
suppose there is $z_j\in M$ with $ \{z_j\}\neq M_i\cap M$ for every
$M_i,\  M_i\in {\mathcal M}$.
\\
  Hence
if $z_j\in M_i$ then $ |M_i\cap M|\geq 2)$ for every $M_i, \
M_i\in{\mathcal M}.$
\\
This means that $M\setminus\{z_j\}$ is a representative system for
${\mathcal M}.$ A contradiction. \fbx
\begin{t00}\rm\label{t3}
Let $Y=\{x_1,\ldots,x_k\}\notin Sep(t,\A).$
 If $Z=\{x_{k+1},\ldots,x_{m}\},\ k< m$ is
a representative system for $Dis(Y,t,\A)$ then $Y\cup Z\in Sep(t,\A).$
\end{t00}
\Pr\rm We will consider the non-trivial case $|Y|\geq 2.$ Clearly
$Dis(Y,t,\A)\neq\emptyset.$
\\
Let us set $V=\{x_{m+1},\ldots,x_n\}=Ess(t,\A)\setminus(Y\cup Z).$
Since, $Z$ is representative system for $Dis(Y,t,\A)$ it follows that
$V_1\notin Dis(Y,t,\A)$
for each $V_1\subseteq V$ and
there is an assignment $\gamma\in Ass(V,\F_0)$ such that
$Y\in Ess(\gamma(t),\A).$
\\
We have to prove that $Z\subset Ess(\gamma(t),\A).$ Suppose this is false.
Without loss of generality
assume that $x_{k+1}\notin Ess(\gamma(t),\A).$
\\
Let $Z_1=\{x_{k+1},x_{j_1},\ldots,x_{j_l}\},\ j_l\leq n$ be a
distributor of $Y$ for $t$ and $\A$ such that $Z_1\cap
Z=\{x_{k+1}\}.$ The existence of $Z_1$ follows by Lemma \ref{l8}.
Thus we have
\\
$\{x_{j_1},\ldots,x_{j_l}\}\subseteq V,$\quad
$Ess(\gamma(t),\A)\cap\{x_{j_1},\ldots,x_{j_l}\}=\emptyset$
and $Ess(\gamma(t),\A)\cap Z_1=\emptyset.$
\\
Let $f_0\in\F_0$ be an arbitrary nullary operation symbol and
$\gamma_1\in Ass(Z_1,\F_0)$ be an assignment defined as follows:
\begin{center}
$\gamma_1(x)=\left\{ \begin{array}{ll}
       f_0 \quad & \mbox{\rm if}\quad  x=x_{k+1};\\
      \gamma(x) \quad & \mbox{\rm if}\quad  x\in Z_1\cap V.
      \end{array}
      \right.
$
\end{center}
Since $(Z_1\setminus V)\cap Ess(\gamma(t),\A)=\emptyset$  it
follows that $Ess(\gamma_1(t),\A)=Ess(\gamma(t),\A).$ Consequently
$Y\subset Ess(\gamma_1(t),\A)$ and $Z_1\notin Dis(Y,t,\A).$ This
is a contradiction. \fbx
\\
There are examples showing that
any representative system $Z$ of the family of distributive sets of $Y$ is a
maximal set for which  $Y\cup Z\in Sep(t,\A)$
i.e. the Theorem \ref{t3} can not be generalized in this direction.

\end{document}